\documentclass{PoS}

\title{Exclusive pi+ pi- production at 7TeV}

\ShortTitle{Exclusive pi+ pi- production at 7TeV}

\author{\speaker{Mohsen Khakzad}%
       \thanks{On behalf of the CMS Collaboration.}\\
      School of Particles and Accelerators, Institute for Research in Fundamental Sciences (IPM), Tehran, Iran \\
      E-mail: \email{mohsen.khakzad@cern.ch}}

\abstract{
We report a measurement of the exclusive production of pairs of charged pions in proton-proton collisions,
dominated by the process ${\rm {pp}} \rightarrow {\rm p}^{(*)} \pi^{+}\pi^{-} {\rm p}^{(*)}$, where $\rm p^{(*)}$ 
stands for a diffractively dissociated proton, the $\pi^{+} \pi^{-}$ pair is emitted at central 
rapidities $y$, and the incident protons stay intact or dissociate without detection ${\rm p^{(*)}}$. 
The measurement is performed with the CMS detector at the LHC, using a data sample corresponding to an integrated luminosity 
of 450~$\mu$b$^{-1}$ collected at a center-of-mass energy of $\sqrt{s}$ = 7 TeV in 2010. 
The cross section measured in the phase space defined by pion transverse momentum $p_{\rm T}>0.2$~GeV/c and rapidity $|y|< 2 $ is 
found to be $20.5~\pm~0.3~(\rm {stat})~(\pm~3.1~\rm {syst})~\pm$~0.8~(lumi)~$\mu$b~\cite{cep_mohsen}.
The differential cross sections for $\pi^{+}\pi^{-}$ pairs 
as a function of the pion pair invariant mass, $p_{\rm T}$, and $y$, as well as a single-pion differential 
cross section as a function of pion $p_{\rm T}$ are also measured and compared to several phenomenological predictions. }
\FullConference{XXIV International Workshop on Deep-Inelastic Scattering and Related Subjects\\
		11-15 April, 2016\\
		DESY Hamburg, Germany}
\begin{document}
\newcommand{\pom}{{\ensuremath{\mathbb{P}}}}
\section{Introduction}
\label{sec:intro}
The process ${\rm pp} \rightarrow {\rm p}^{(*)}\pi^+\pi^-{\rm p}^{(*)}$ is one in which two pions are produced and the colliding protons remain intact (exclusive) 
or dissociate into low-mass systems (semi-exclusive).  At high center-of-mass energies, this process is expected to 
be dominated by double pomeron exchange (DPE)~\cite{cep_mass} as depicted in Fig. 1a. A smaller contribution can occur 
through $\rho$ meson photoproduction as illustrated in Fig 1b.  The contribution from the two-photon fusion process 
is expected to have a much smaller cross section and is therefore not considered.
\begin{figure}[htbp]
\centerline{
  \mbox{\includegraphics[width=3.1in]{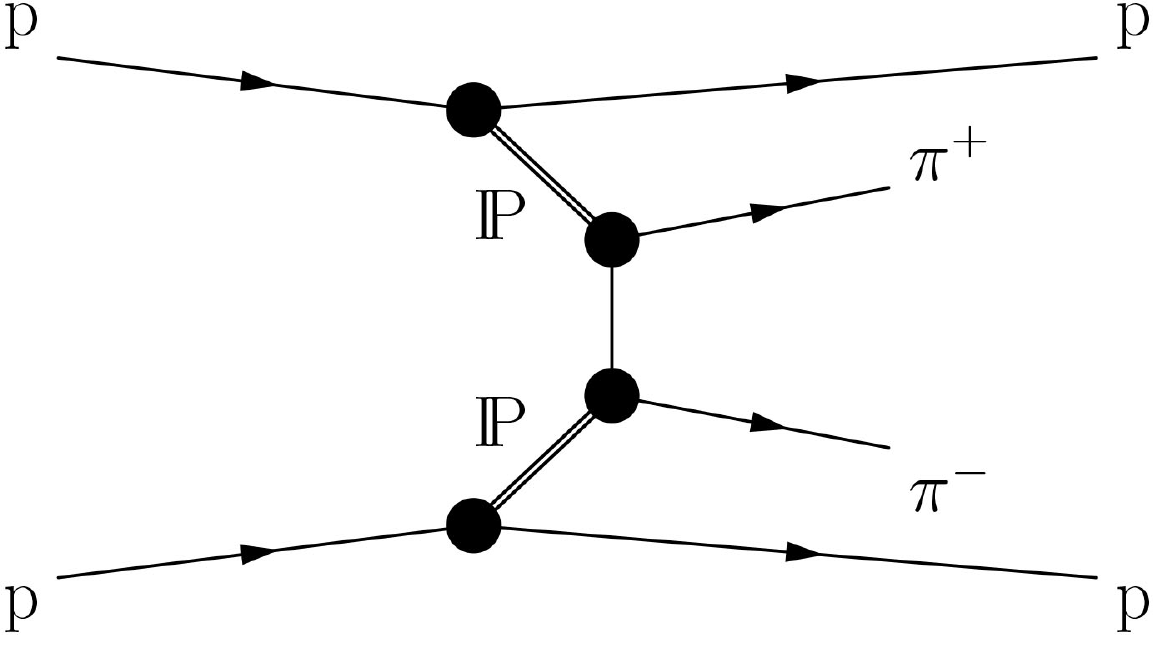}}
  \put(-115, -1){\Large a) }	
  \put(110, -1){\Large b) }
   \hspace{0.1in}	
 \mbox{\includegraphics[width=3.1in]{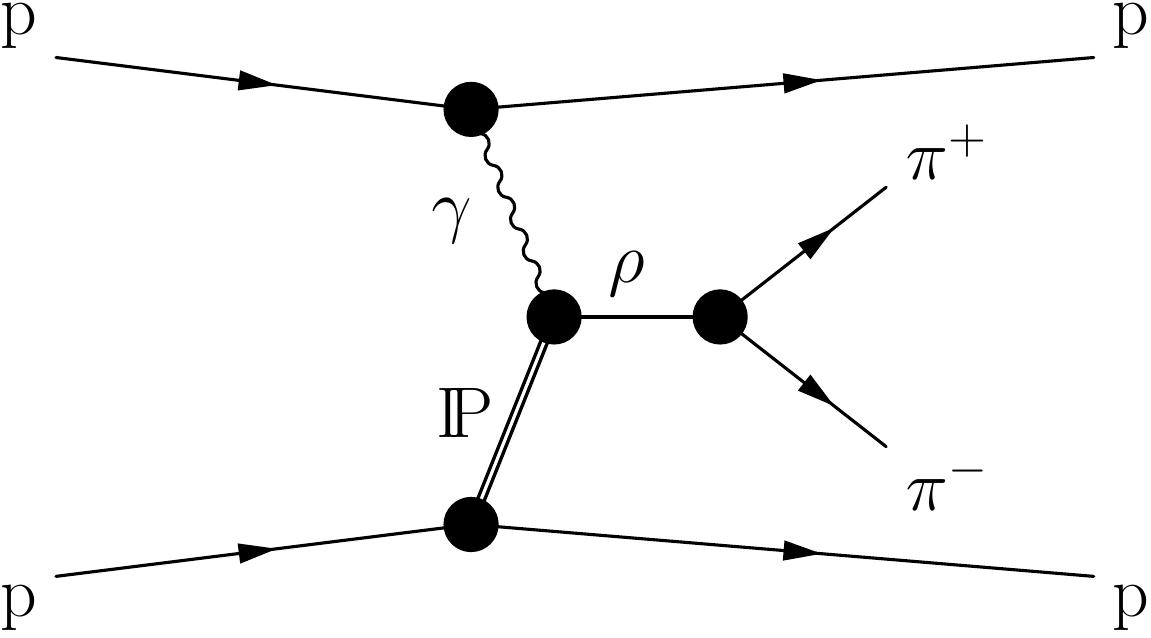}} }
  \caption{(a) A double pomeron exchange (DPE) diagram that produces a central $\pi^{+}\pi^{-}$ system in
    proton-proton collisions, and  
  (b) a photon-pomeron interaction that produces an exclusively produced $\rho$~meson that subsequently decays into a pair of pions.}
  \label{cep}
  \end{figure}
DPE process is useful for meson spectroscopy as the central state has a limited set of quantum 
numbers (isospin of 0, even spin, positive parity, and positive charge-conjugation parity) that can 
be used as a quantum-number filter, with selection rules differing from those of other production processes like 
${\rm {e}^+ {\rm e}^-} \rightarrow \pi^+\pi^-$. Pomeron exchange at low $p_{\rm T}$
is dominated by two or more gluons, and DPE is therefore considered a good channel for the production of 
glueballs~\cite{cep_mass, cep_ex_1, cep_63_2}. In addition, the DPE may provide a better understanding of the physics of pomeron exchange, which is 
a nonperturbative process, and is consequently model-dependent.

A variety of exclusive final states in hadron-hadron collisions have been investigated, starting from the measurements in pp collisions 
at $\sqrt{s}$ = 23 to 63 GeV at the CERN ISR~\cite{cep_ex_1, cep_63_2}. 
Exclusive production of hadrons at central rapidities is usually described in terms of DPE processes, when the
mass of the central system is not very large (below 3 $GeV/c^2$), or perturbatively, 
as in central exclusive-production approaches, where the partonic structure of the pomeron 
(color singlet exchange with vacuum quantum numbers) is explicitly taken into account. 
At $\sqrt{s}$ = 63 GeV, DPE was observed in the exclusive $\pi^{+} \pi^{-}$ channel, while at lower (fixed-target) energies,   
exclusive two-hadron production showed large additional contributions from Reggeon exchange. 

This paper presents a measurement of exclusive $\pi^{+} \pi^{-}$  production in the nonperturbative 
regime up to invariant masses ${\rm M}({\pi \pi})$ $\approx$ 3 $\rm {GeV/c}^2$ in pp collisions at $\sqrt{s}$ = 7 TeV. 
The integrated and differential cross sections are reported for exclusive and semiexclusive pair production in the phase space of 
pion transverse momentum $p_{\rm T}>$~0.2~GeV/c and rapidity $|y|<$~2. 

\section{The CMS detector}
A detailed description of the CMS detector, together with a definition of the coordinate system used 
and the relevant kinematic variables, can be found in Ref.~\cite{cms_detector}. 
The central feature of the CMS apparatus is a superconducting solenoid, 13~m in length and 6~m 
in diameter, providing an axial magnetic field of 3.8 T. 
Silicon pixel and strip trackers are surrounded by a crystal electromagnetic calorimeter (ECAL) and 
a brass and scintillator hadron calorimeter (HCAL). 
Muons are measured in gas-ionization chambers embedded in the steel flux-return yoke of the magnet.

Particles created in collisions in the center of the detector first traverse the tracker, a system of silicon sensors designed to provide a precise and efficient measurement 
of the trajectories of charged particles.The inner tracker measures charged particles within the pseudorapidity range $|\eta| <$ 2.5.
For isolated pions with $p_{\rm T}$=1 GeV/c and $|\eta|<1.4$, the transverse (longitudinal) impact parameter 
resolution is about 90 (100--200) microns and the $p_{\rm T}$ resolution varies from 0.8 to 2\%.
The overall length of the tracker is 5.4 m, and its outer diameter is 2.4 m. 
The beam pickup for timing (BPTX) devices are used to trigger the detector readout. 
They are located around the beam pipe at a distance of 175 m from the interaction point (IP) on 
either side, and are designed to provide precise information on the LHC bunch structure and the timing of the incoming beams.

The ECAL provides coverage in the range of  $|\eta| <$~1.5 in the barrel region (EB) and 1.5~$< |\eta| <$~3.0 in the two endcap regions (EE).
In the region $|\eta| < 1.74$, the HCAL cells have widths of 0.087 in pseudorapidity and 0.087 in azimuth ($\phi$). 
In the $\eta$-$\phi$ plane, and for $|\eta| < 1.48$, the HCAL cells map on to $5 \times 5$ ECAL crystals arrays to form calorimeter towers projecting radially 
outwards from close to the nominal interaction point. At larger values of $|\eta|$, the size of the towers increases and the matching ECAL arrays contain fewer crystals.
Two forward calorimeters (HF) cover 2.9~$< |\eta| <$~4.9.
\section{Data analysis}
\subsection{\label{event_select} Event selection}
The data used in this analysis correspond to an integrated luminosity of 450~$\mu$b$^{-1}$ 
and were collected in 2010 at $\sqrt{s}$ = 7 TeV at a low instantaneous luminosity, with $\approx$ 1 pp interaction per bunch crossing. 
Events were selected using a ``zero-bias'' trigger, which requires only the presence of crossing proton bunches at the CMS interaction point. 
This trigger was prescaled by a factor that varied from 200 to 33\,000 to control the output rate. The prescale factor of the trigger is taken into 
account in the calculation of the integrated luminosity used to extract the cross section. 
The trigger is provided by the BPTX detectors that provide beam bunch detection with $\approx$ 100\% efficiency. 

Events are required to have two charged tracks coming from a common point on the beam line, 
with no additional tracks and no activity in the calorimeter above the noise thresholds 
to reject nonexclusive events as well as events with more than a single interaction vertex from beam (pileup). 

The following requirements are imposed to select well-reconstructed tracks consistent with those originating from a
pp collision: 

\begin{itemize}
\item A standard high-purity selection is applied to reduce the number of false tracks.  
This selection uses information on the number of hits, normalized $\chi^2$, and transverse and longitudinal impact parameters for the track.
\item The two selected tracks are required to intersect at a vertex within $|z_{\rm vtx}| < $15 cm of the longitudinal center of the detector. 
\end{itemize}
A kinematic region with high tracking efficiency, that is also well separated from the edges of the tracker
acceptance, is defined by requiring each particle to have $p_{\rm T} >$~0.2~GeV/c and $|y| < 2$, where $y$ is calculated assuming it to be a pion. 
Hadron identification using specific ionization (dE/dx) shows that at low momenta, where $\pi$ and $K$ mesons can be
distinguished, more than 92.7\% of the tracks are due to pions. Since there is no distinct $\pi/K$ separation above approximately 0.6 GeV/c, 
we do not attempt to subtract the $K^+K^-$ background.
At this stage of the selection, there is no requirement on the charge of the two tracks. Since exclusive production of  
same-sign pairs ${\rm p p \rightarrow p}  (\pi^\pm \pi^\pm)  {\rm p}$ is forbidden by charge conservation, same-sign events  
can be used as a control sample to study residual multihadron backgrounds, where 
two or more charged particles are not detected or do not pass the reconstruction criteria.

In addition to having only two charged-particle tracks in the event (apart from the undetected ${\rm p^{(*)}}$),
exclusive $\pi^{+} \pi^{-}$ events are required to have no 
extra activity in the calorimeters up to $|\eta|$~=~4.9. 
This condition is applied by counting the number $N_{\rm {extra}}$ of calorimeter towers that contain signals above 
the noise thresholds cones with $\Delta R >$~0.1 around the charged tracks, where $\Delta R$ = $\sqrt{(\Delta\eta^2) + (\Delta\phi^2)}$. 
The applied noise thresholds are determined for each region of the HCAL, ECAL and HF, using samples triggered on unpaired proton beam bunches.  
The threshold energies range from $\sum E$ = 0.52~GeV in the EB to 4.0~GeV 
in the HF, where $\sum E$ is the sum of the calorimeter tower energies. 

\begin{figure}[htbp]
\centerline{
\includegraphics[width=3.1in]{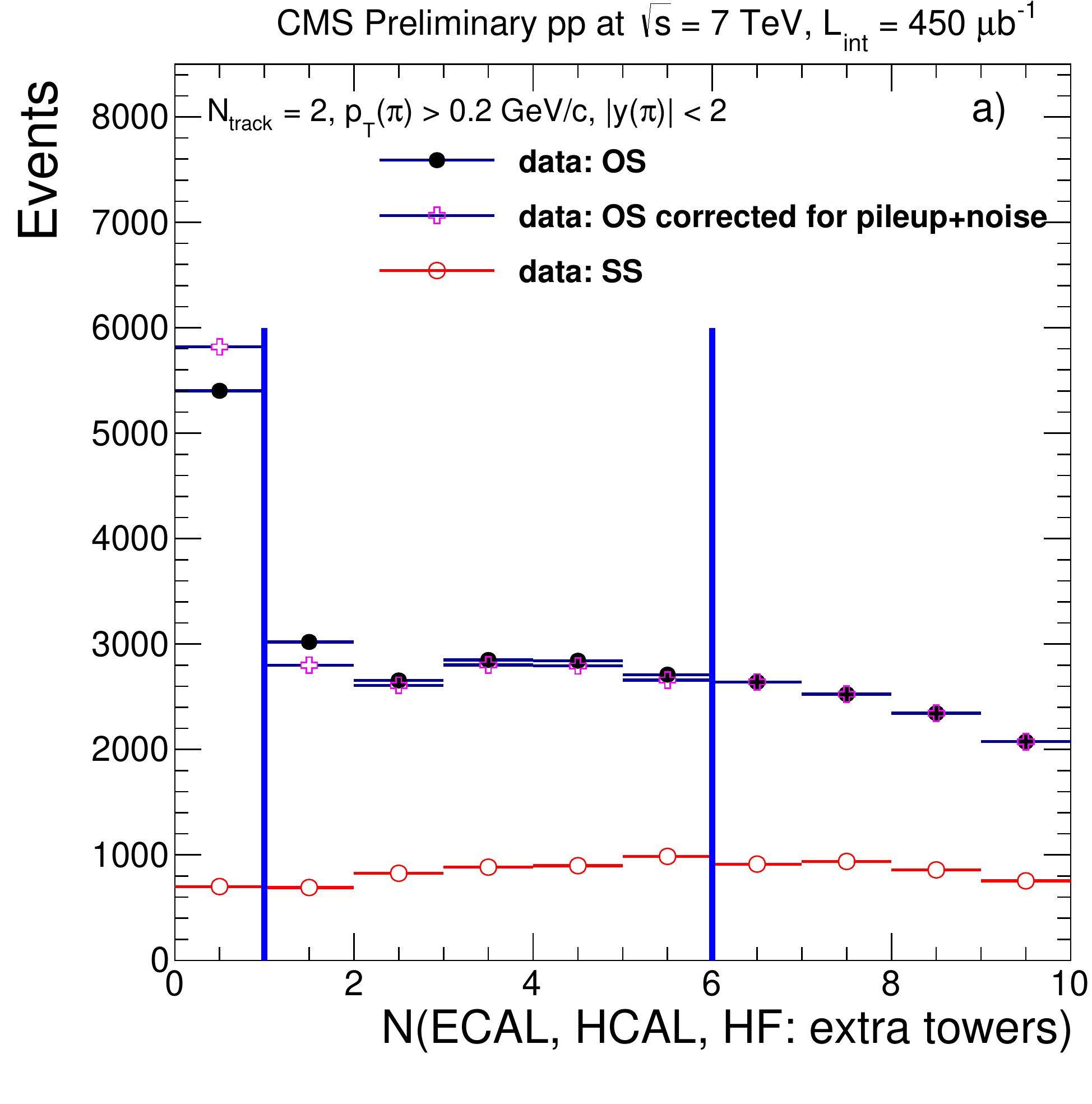}
\includegraphics[width=3.1in]{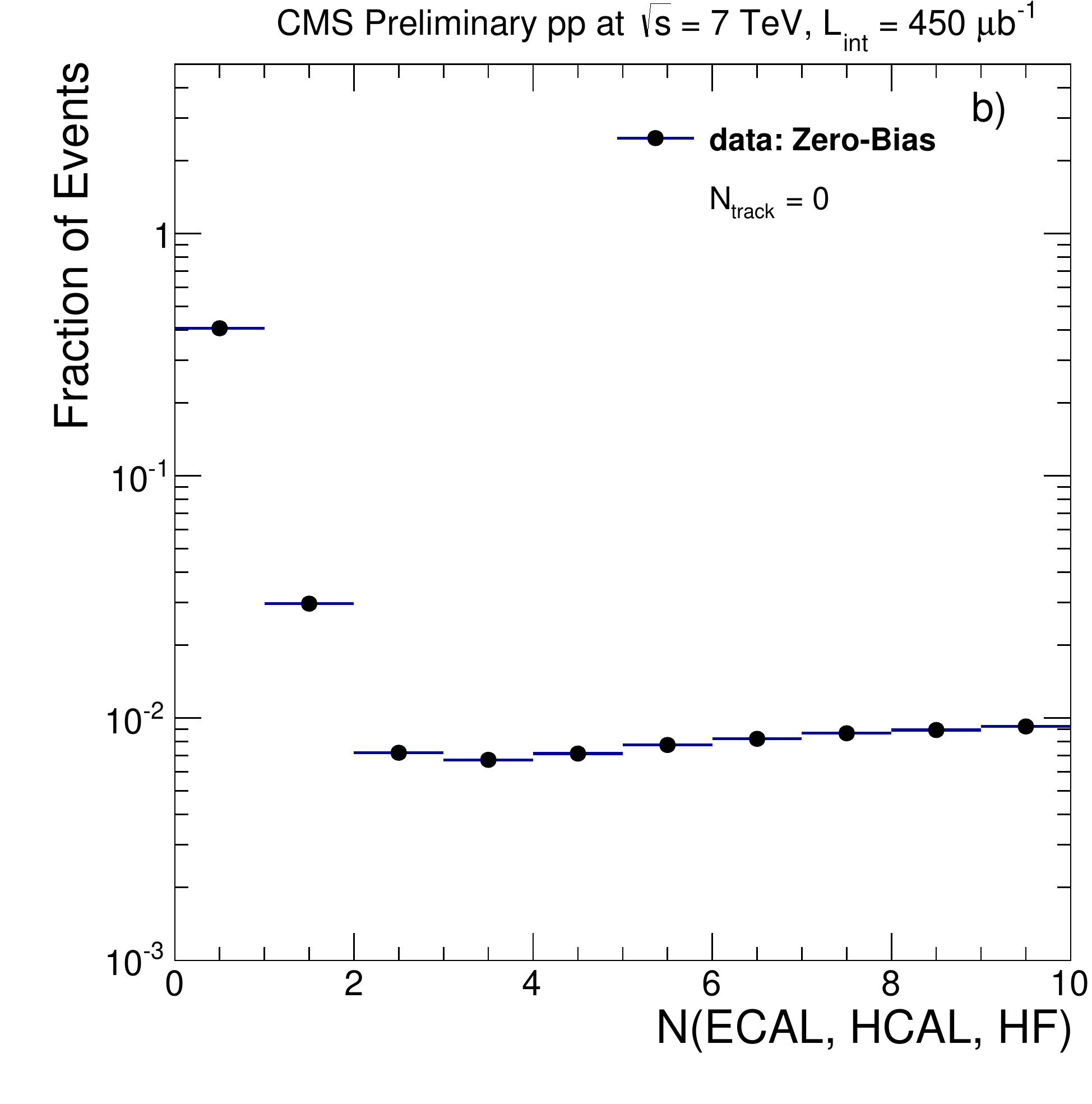} }
\caption{(a) Distribution of ECAL+HCAL tower multiplicities above noise threshold in events with colliding beams 
and pairs of reconstructed opposite-sign (OS) and same-sign (SS) tracks. The distribution after the pileup correction 
is shown by crosses. The vertical lines indicate the background control region. (b) Distribution of ECAL+HCAL tower 
multiplicities above noise threshold in the zero-bias sample with no tracks required, normalized to unity. }
\label{signal}
\end{figure}

Figure~\ref{signal}a shows $N_{\rm extra}$ in the selected two-track events of opposite-sign (OS) 
pairs and for same-sign (SS) pairs. A clear excess is observed in OS events with $N_{\rm extra}$~=~0 (``signal region") compared to 
events with one or more extra towers. In contrast, no excess of events with $N_{\rm extra}$~=~0 towers is observed in the SS sample.
Figure~\ref{signal}b shows the distribution in the number of towers (normalized to 1.0) above threshold in zero-bias  
bunch-crossing events with no tracks, a sample that has almost no 
inelastic interactions. While 40\% of these events have no towers above threshold, the next bin (``1-tower") also shows an excess above
the continuous distribution (considered to be background noise). While we select the events in the $N_{\rm extra}$~=~0 bin as the exclusive $\pi^+\pi^-$
candidates, we correct the cross section for migration to the $N_{\rm extra}$ = 1 bin. The nonexclusive background is estimated from the bins $N_{\rm extra}$ = 2$-$5.

The result is given by the open crosses in Fig.~\ref{signal}a. 
There is a 7.3\% migration from the $N_{\rm extra}$ = 1 bin to the signal bin $N_{\rm extra}$~=~0. The bins $N_{\rm extra}$ = 2$-$5 
contribute an additional 1.8\% to the signal bin. Before and after correcting for migration, there are 5402 and 5818 signal events, 
respectively. The latter are used to extract the cross section for exclusive $\pi^+\pi^-$ production.
For an integrated luminosity of 450~$\mu$b$^{-1}$, 5402 OS events are observed with no extra towers 
above threshold, while there are 700 SS events passing all  selection criteria. Within the SS sample, there is no significant charge asymmetry observed, with  
368 negatively-charged pairs and 332 positively-charged pairs. 
\section{\label{cross-section}Cross Sections definition and results}
\begin{figure}[htbp]
\centerline{
  \includegraphics[width=3.1in]{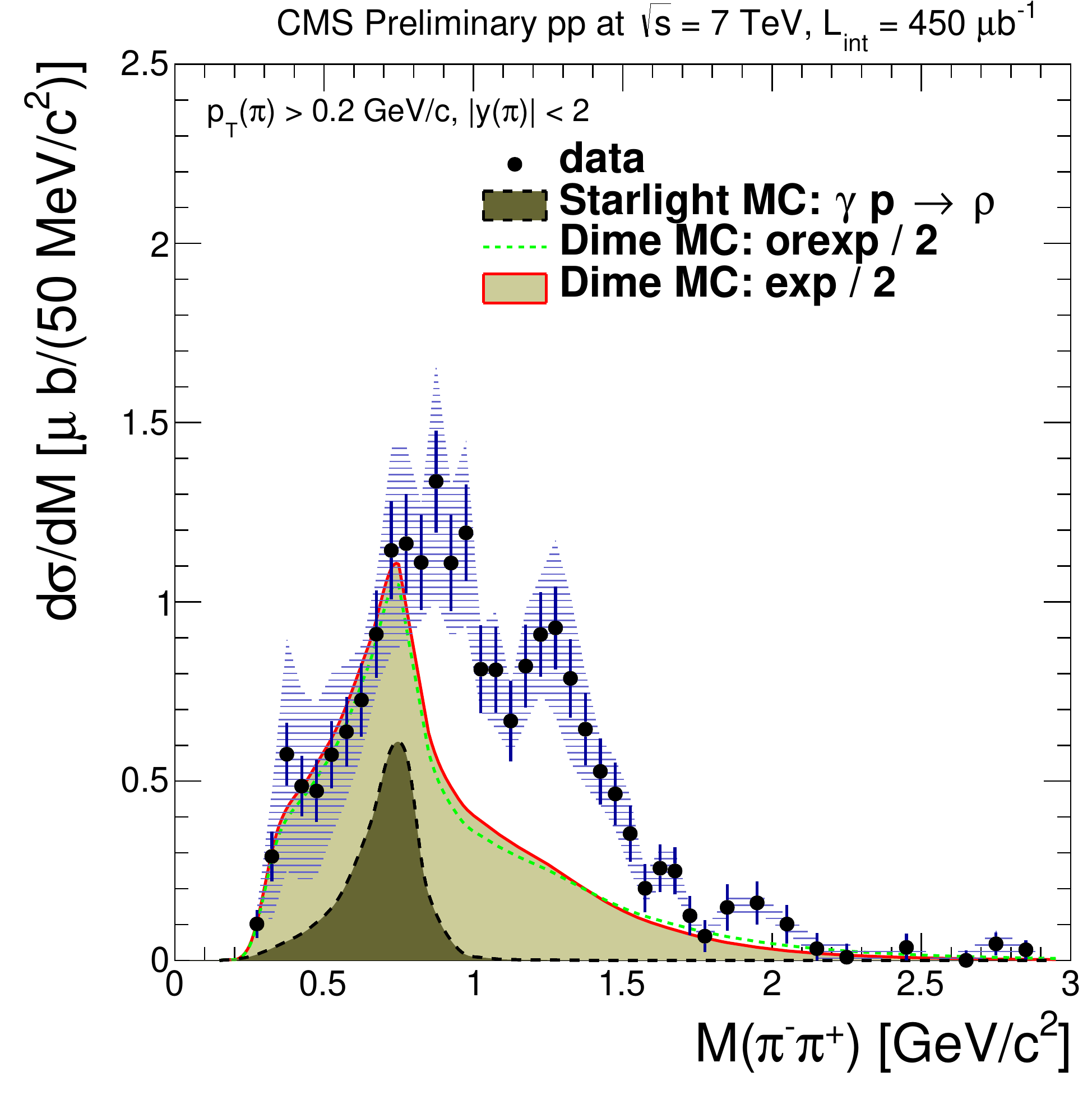}
  \includegraphics[width=3.1in]{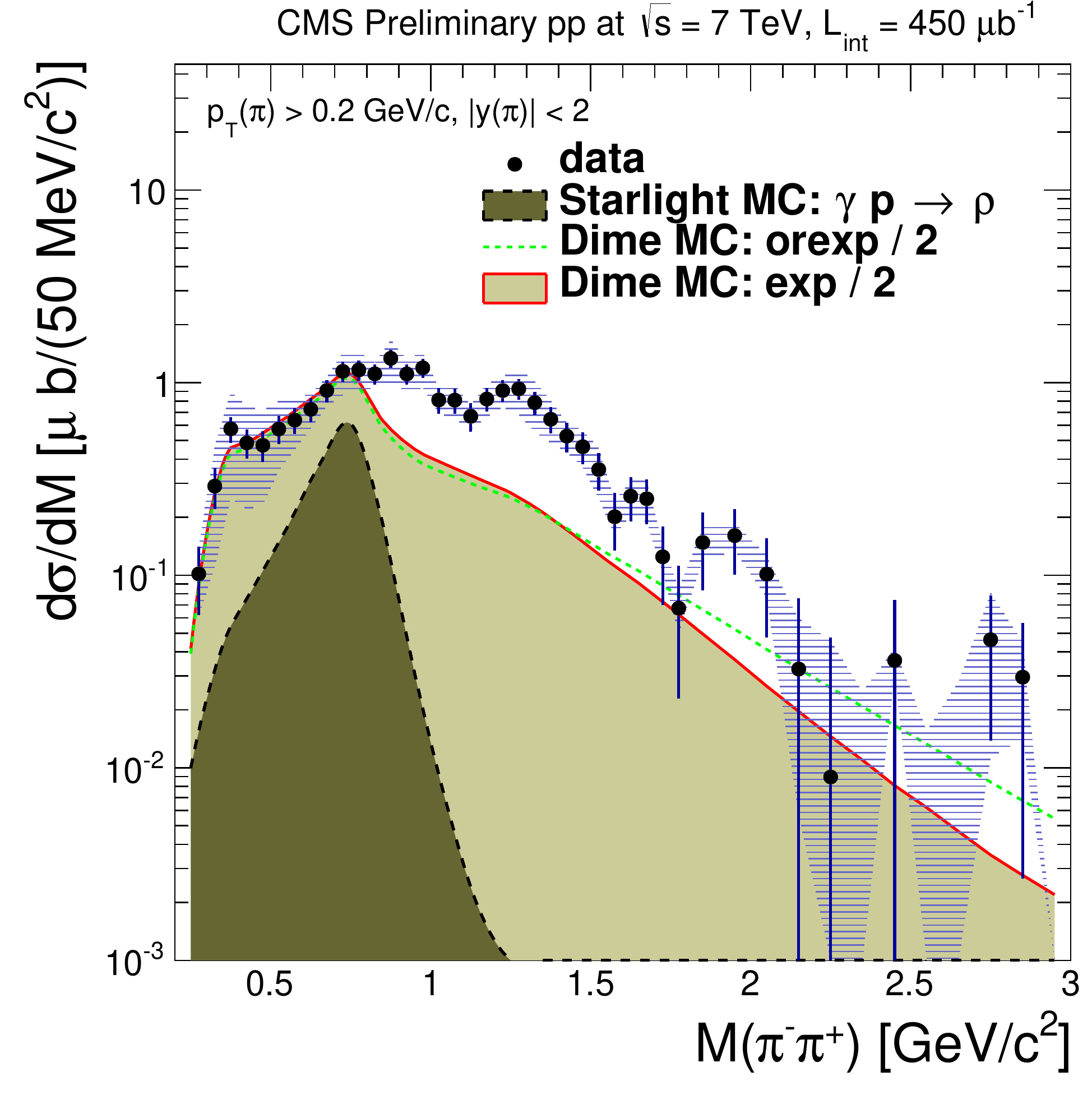} }
\caption{Corrected differential cross sections for $p p \rightarrow p^{(*)} \pi^{+}\pi^{-} p^{(*)}$ as a function dipion invariant-mass, 
compared to the stacked predictions of DPE 
production from the DIME MC (solid curve) and  of $\rho$ photoproduction from STARLIGHT (dashed curve). The shaded band shows 
the overall systematic uncertainty, and the error bar indicates the statistical uncertainty. The results are plotted on a linear scale (a) and a 
logarithmic scale (b). } 
\label{dSigma_dM}
\end{figure}

The ${\rm pp \rightarrow p^{(*)} } \pi^{+}\pi^{-}  {\rm p^{(*)}}$ cross section is obtained from the number of OS events passing all 
selection criteria and corrected for pileup and noise, with the background subtracted.
In addition, a correction factor is applied to account for the fraction of triggered events that contain only a single pp collision.
The fully corrected differential exclusive $\pi^{+}\pi^{-}$ cross sections after background subtraction as a 
function of the pion pair invariant mass ${\rm M}(\pi\pi)$ is shown in Fig.~\ref{dSigma_dM}. 
The error bars on the data points show the statistical uncertainty, while the systematic uncertainties are shown by the shaded band. 
The DIME MC predictions~\cite{durham}, with two form-factors, exponential and Orear labelled by "exp" and "Orear", respectively, 
for exclusive $\pi^{+}\pi^{-}$ production without proton dissociation, and the STARLIGHT predictions~\cite{starlight}. 
for $\rho$ photoproduction are shown for comparison. 
In the {\sc Dime MC} model, the effect of so-called "enhanced" rescattering (the possibility that the exchanged pomeron from one proton may 
interact with the other proton and produce additional particles) is not included. 
However, this is expected to give a factor of two suppression in the cross section. 
Therefore, the {\sc Dime MC} predictions have been divided by two to account for this effect. 
The sum of the predicted cross sections is $\approx$12~$\mu$b, compared to the measured value of 
20.5~$\pm$~0.3~(stat)~$\pm$~3.1~(syst)~$\pm$~0.8~(lumi)~$\mu$b. 
It should be noted that neither the DIME nor the STARLIGHT predictions 
include the effect of low-mass proton dissociation, which would increase the 
visible cross section. 
\section{Summary}
Integrated and differential cross sections for charged pion-pair production in the exclusive or semiexclusive reactions, 
${\rm p p \rightarrow p^{(*)} } \pi^{+} \pi^{-} {\rm p^{(*)}}$, 
where the incident protons stay intact or dissociate into undetected low-mass states, have been measured in proton-proton
collisions at $\sqrt{s} = 7$ TeV  with the CMS detector at the LHC using 
a data sample corresponding to an integrated luminosity of 450~$\mu$b$^{-1}$.  
The integrated cross section for the sum of the exclusive and semiexclusive contributions to the region of 
 space of individual pions $p_{\rm T} >$~0.2~GeV/c and $|y| <$~2, with no additional particles above threshold, is found to be 
20.5~$\pm$ 0.3~(stat)~$\pm$ 3.1~(syst)~$\pm$~0.8~(lumi)~$\mu$b. The results shown here are preliminary. 


%
\end{document}